\documentclass[prl,showpacs,preprintnumbers,twocolumn,10pt]{revtex4}%
\usepackage{amsmath}
\usepackage{graphicx}
\usepackage{amsfonts}
\usepackage{amssymb}

\begin{document}
\title{Intermittency in an Optomechanical Cavity Near a Subcritical Hopf Bifurcation}
\author{Oren Suchoi}
\author{Lior Ella}
\author{Oleg Shtempluk}
\author{Eyal Buks}
\affiliation{Department of electrical engineering, Technion, Haifa 32000, Israel}
\date{\today }

\begin{abstract}
We experimentally study an optomechanical cavity consisting of an oscillating
mechanical resonator embedded in a superconducting microwave transmission line
cavity. Tunable optomechanical coupling between the mechanical resonator and
the microwave cavity is introduced by positioning a niobium-coated single mode
optical fiber above the mechanical resonator. The capacitance between the
mechanical resonator and the coated fiber gives rise to optomechanical
coupling, which can be controlled by varying the fiber-resonator distance. We
study radiation pressure induced self-excited oscillations as a function of
microwave driving parameters (frequency and power). Intermittency between
limit cycle and steady state behaviors is observed with blue-detuned driving
frequency. The experimental results are accounted for by a model that takes
into account the Duffing-like nonlinearity of the microwave cavity. A
stability analysis reveals a subcritical Hopf bifurcation near the region
where intermittency is observed.

\end{abstract}

\pacs{46.40.- f, 05.45.- a, 65.40.De, 62.40.+ i}
\maketitle

The field of cavity optomechanics
\cite{Braginsky_653,Marquardt_0905_0566,girvin_Trend,Kippenberg_1172} deals
with a family of systems, each is composed of two coupled elements. The first
one is a mechanical resonator, commonly having low damping rate, and the
second one is an electromagnetic cavity, which is typically externally driven.
Both radiation pressure
\cite{Corbitt_150802,Gigan_67,Arcizet_et_al_06,Schliesser_243905,Thompson_72,Nichols_307,Carmon_et_al_05,Jayich_et_al_08,Schliesser_et_al_08}
and bolometric force
\cite{Metzger_1002,Jourdan_et_al_08,Metzger_133903,Marino&Marin_10,Restrepo_860,Liberato_et_al_10,Marquardt_103901,Paternostro_et_al_06,Aubin_1018,Marquardt_103901,Zaitsev_046605,Zaitsev_1589}
can give rise to the coupling between the mechanical resonator and the cavity.
In recent years a variety of cavity optomechanical systems have been
constructed and studied
\cite{Hane_179,Kippenberg_1172,Corbitt_S675,Corbitt_150802,Metzger_1002,Gigan_67,Arcizet_et_al_06,Kleckner_75,Favero_104101,Regal_555,Carmon_223902,Schliesser_243905,Thompson_72,Chan_89,Teufel_359,Teufel_204,OConnell_697,Groeblacher_181104}%
, and phenomena such as mode cooling
\cite{Teufel_359,Teufel_204,Groblacher_485,Schliesser_509,Chan_89},
self-excited oscillations
\cite{Hane_179,Kim_1454225,Aubin_1018,Carmon_223902,Marquardt_103901,Corbitt_021802,Carmon_123901,Metzger_133903,Regal_555}
and optically induced transparency
\cite{Weis_1520,Karuza_013804,Safavi-Naeini_69,Ojanen_1402_6929} have been
investigated. In addition to applications in metrology
\cite{Braginsky_QM,Clerk_1155} and photonics
\cite{Hossein_Zadeh_276,Bagheri_726,Zhou_179}, the appeal of optomechanics
lies in the potential for observation of macroscopic quantum behavior in
mechanical systems
\cite{Teufel_204,OConnell_697,Kimble_et_al_01,Genes_et_al_08,Teufel_et_al_10,Rodrigues_053601,Qian_1112_6200,Palomaki_710,He_052121,Pikovski_393,Poot_273,Meaney_056202,Walter_1307_7044,Lorch_011015,Galland_1312_4303,Kiesewetter_1312_6474,Bahrami_1402_5421,Farace_1306_1142,Xu_063819,Weinstein_1404_3242,Xu_1404_3726}%
. While much experimental and theoretical progress has been made in reaching
the ground state and observing the linear dynamics of mechanical objects, it
is becoming appreciated that nonlinearity allows the creation of non-classical
mechanical states \cite{Loerch_11015} and can be exploited for improving the
efficiency of optomechanical cooling \cite{Nation_104516}. It is therefore
important to study and shed light on the nonlinear dynamics of these devices.%

\begin{figure}

%\begin{center}
\includegraphics[width=0.5\textwidth
%natheight=23.947500in,
%natwidth=18.541599in,
%height=4.9113in,
%width=3.8086in
]{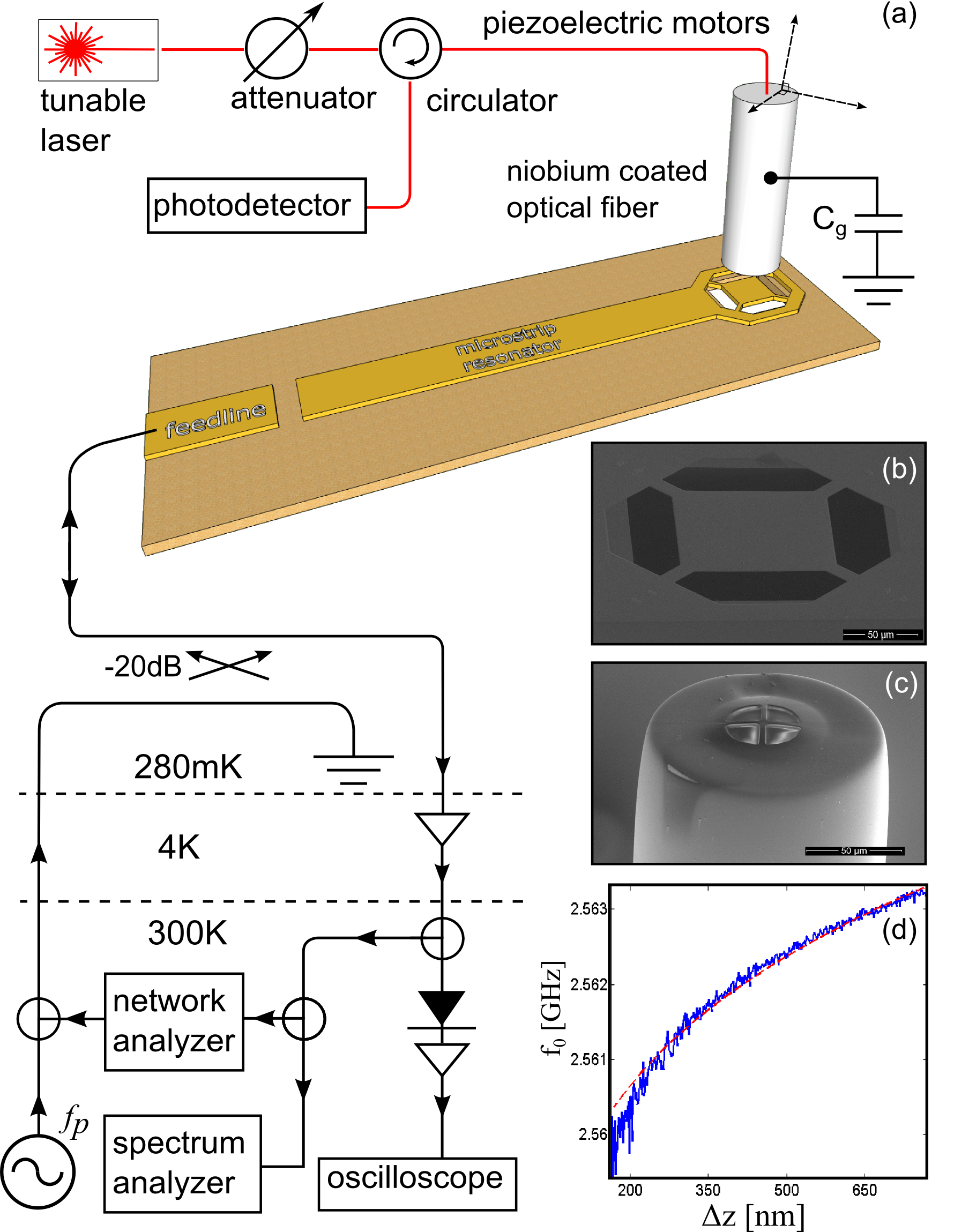}
\
\caption{{}Experimental Setup (color online). The Microwave cavity is a
microstrip made of aluminum over high resistivity silicon wafer coated with a
$100\operatorname{nm}$ thick SiN layer. The mechanical resonator at the end of
the microstrip is a suspended trampoline supported by four beams. Electron
micrograph of the trampoline is shown in panel (b). The optomechanical
coupling is generated using a niobium coated optical fiber that is positioned at
sub-micron distance from the trampoline. Several windows are opened in the niobium
layer on the fiber tip using FIB, as can be seen in panel (c). The optical
setup (seen above the sample) allows using the optical fiber for displacement
detection, whereas the microwave setup (seen below the sample) is employed for
measuring the cavity response. The coated fiber is galvanically connected to
both AC and DC voltage sources with a bias-T (not seen in the sketch), which
can be used to externally actuate the mechanical resonator. The sample is
mounted inside a closed copper package, which is internally coated with niobium.
Measurements are performed in a dilution refrigerator operating at a
temperature of $0.28\operatorname{K}$ and in vacuum. Panel (d) shows the
measured (solid) and calculated (dashed) cavity resonance frequency $f_{0}$
vs. fiber-trampoline distance $\Delta z$.}%
\label{Setup_fig}
%\end{center}
\end{figure}

In this work we experimentally study self-excited oscillations in an
optomechanical cavity operating in the microwave band. We introduce a novel
method for achieving strong and tunable optomechanical coupling, which is
based on positioning a metallically coated optical fiber near the mechanical
resonator. The microwave cavity, which is made of a superconducting aluminum
microstrip, exhibits Kerr type nonlinearity
\cite{Dahm_2002,Suchoi_174525,Yurke_5054,Boerkje_53603}, which significantly
affects the dynamics of the entire optomechanical system \cite{Nation_104516}.
We study the dependence of the self-excited oscillations on the driving
parameters of the cavity and found that a good agreement with theory can be
obtained provided that cavity nonlinearity is taken into account
\cite{Nation_104516}. We experimentally find that in a certain region of drive
parameters the system exhibits random jumps between a limit-cycle (i.e.
self-excited oscillations) and a steady-state. A theoretical stability
analysis reveals that this observed intermittency behavior occurs near a
subcritical \cite{Larson_021804,Holmes_066203,Blocher_119,Meaney_056202} Hopf
bifurcation \cite{Walter_1307_7044,Lorch_011015}.

The experimental setup is schematically depicted in Fig. \ref{Setup_fig}.
Magnetron DC sputtering is employed for coating a high resistivity silicon
wafer with aluminum. The aluminum layer is annealed \textit{in situ} at 400$^{\circ}$C for $10-30$ minutes to reduce internal stress in the layer
\cite{Jaecklin_269}. A standard photo-lithography process is used to pattern
the microwave microstrip cavity. At the open end of the microstrip a 100nm
thick SiN membrane is fabricated \cite{Zaitsev_046605}. The mechanical
resonator is made by releasing a $100\times100\mu$m$^{2}$ trampoline supported by four beams using electron cyclotron resonance (ECR) dry etch. At the other end the cavity is weakly coupled to a feedline, which guides both the injected and reflected microwave signals. The results presented here are obtained with a device having a fundamental cavity resonance frequency $\omega_{a}/2\pi=2.5465$ GHz, cavity linear damping rate $\gamma_{a}/2\pi=$420kHz, fundamental mechanical resonance frequency $\omega_{b}/2\pi=$ 12.1kHz and mechanical Q-factor $Q_{b}=3700$.

As can be seen in Fig. \ref{Setup_fig}, a single mode optical fiber coated
with niobium is placed above the suspended trampoline. In the presence of the
coated fiber two optomechanical cavities are formed, one in the microwave band
and the other in the optical band \cite{Zaitsev_46605}. The fact that both
optomechanical cavities share the same mechanical resonator can be exploited
for conversion between microwave and optical photons
\cite{Andrews_1310_5276,Jiang_1404_3928,Fong_1404_3427,Clader_1403_7056,Bochmann_122602}. However, in the present work we employ the optical cavity and the optical
setup seen in Fig. \ref{Setup_fig} only for fiber positioning and for
characterization of the mechanical resonator at high temperatures, whereas all
low temperature measurements that are discussed below are done in the
microwave band only.

We employ a telecom single mode optical fiber having a fiber Bragg grating
(FBG) mirror \cite{Zaitsev_46605} and a focusing lens, made by melting the
fiber tip. Magnetron DC sputtering is used for coating the fiber with niobium. To
allow optical transmission, we etch the niobium coating using focused ion beam
(FIB), exposing thus the core of the fiber at the tip. A cryogenic
piezoelectric 3-axis positioning system having sub-nanometer resolution is
employed for manipulating the position of the optical fiber.

Optomechanical coupling between the microwave cavity and the mechanical
resonator is introduced due to the capacitance between the coated fiber and
the suspended trampoline, which is given by approximately $C_{\mathrm{SP}%
}=2\pi\epsilon_{0}R_{\mathrm{F}}\log\left(  R_{\mathrm{F}}/\Delta z\right)  $
\cite{Russel_10}, where $R_F$ =350 $\mu$m is the radius of curvature of the melted fiber tip, $\Delta z$ is the fiber-trampoline distance and $\epsilon_{0}$ is the vacuum permittivity. A hole of diameter $d_{\mathrm{H}}=$ 2.4mm and depth $h_{\mathrm{H}}= $ 2.7mm is drilled in the sample package, which is made of copper, above the trampoline in order to allow inserting the optical fiber, which has an outer diameter of $d_{\mathrm{F}}=$125 $\mu$m. When the fiber is centered inside the hole, the fiber-package coaxial capacitance is given by $C_{\mathrm{g}}=2\pi\epsilon_{0}h_{\mathrm{H}} /\log\left(  d_{\mathrm{H}}/d_{\mathrm{F}}\right)  $ \cite{Pozar_MWE}. When
radiation loss is disregarded, the effect of the coated fiber on microwave
cavity modes can be accounted for by assuming that a termination having purely
imaginary impedance given by $Z_{\mathrm{T}}=1/i\omega C$, where
$C^{-1}=C_{\mathrm{SP}}^{-1}+C_{\mathrm{g}}^{-1}$, has been introduced between
the microstrip end and ground. The frequencies $f_{a}$ of the cavity modes can
be found by solving $\tan\left(  \kappa l_{\mathrm{M}}\right)  =iZ_{0}%
/Z_{\mathrm{T}}$ \cite{Pozar_MWE}, where the propagation constant $\kappa$ is
related to $f_{a}$ by $f_{a}=\kappa c^{\prime}/2\pi$, $c^{\prime}$ is the
propagation velocity in the microstrip, $l_{\mathrm{M}}$ is the length of the
microstrip, and $Z_{0}=$48 $\Omega$ is its characteristic impedance. Comparison between the measured and calculated values of the cavity fundamental mode frequency $f_{0}$ is seen in panel (d) of Fig. \ref{Setup_fig}. The dependence of $f_{0}$ on
fiber-trampoline distance $\Delta z$ allows the extraction of optomechanical
coupling coefficient $\Omega$, which is found to be given by $\Omega/x_{b0}$=55MHz $\mu$m$^{-1}$ for our chosen operating point, where $x_{b0}=\sqrt{\hbar/2m\omega_{b}%
}=1.1\times10^{-8}\mu$m is the mechanical zero point amplitude, and where $m=5.4\times10^{-12}$kg is the effective mass of the mechanical mode.%

\begin{figure*}

%\begin{center}
\includegraphics[width=\textwidth
%natheight=5.156000in,
%natwidth=9.625400in,
%height=3.4688in,
%width=6.4524in
]
{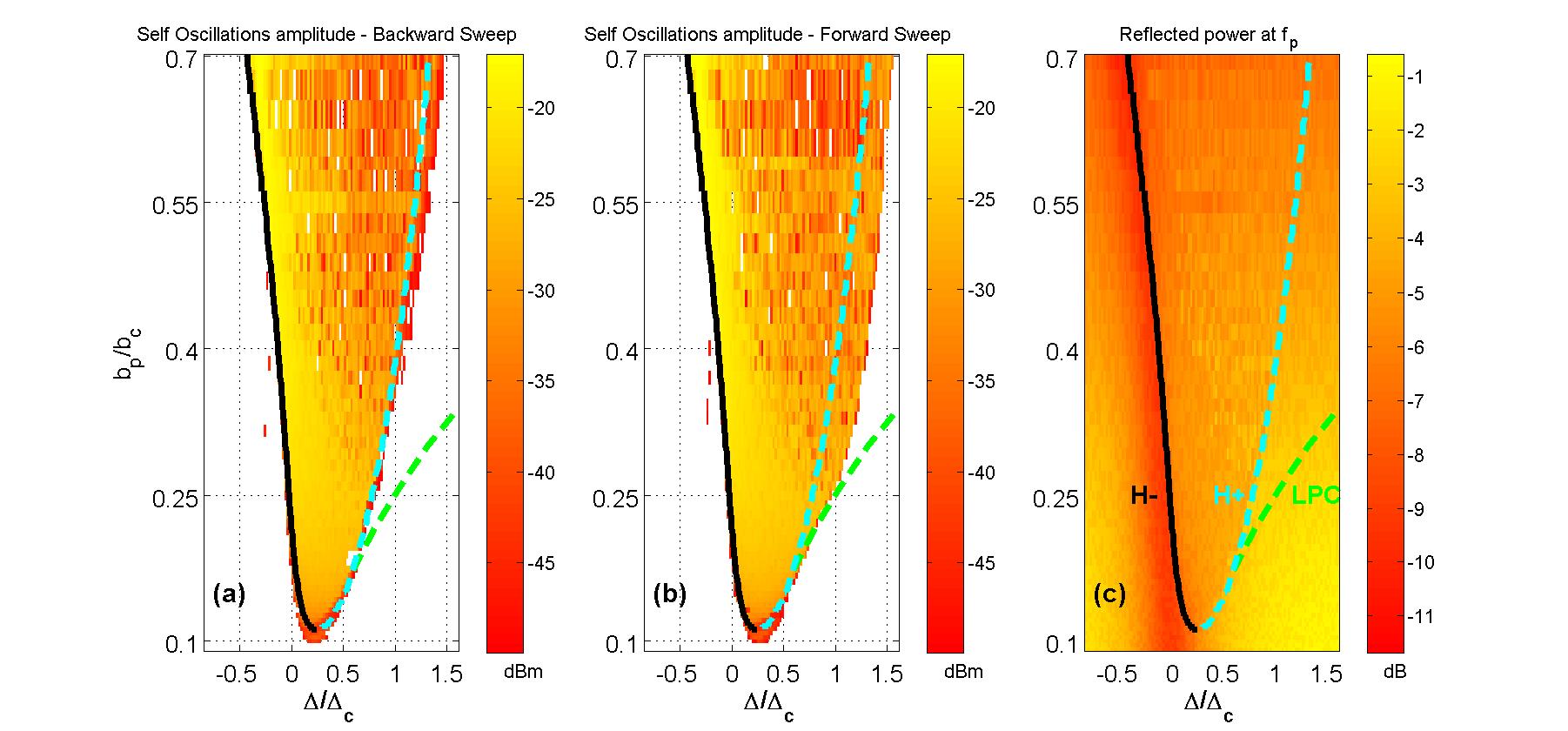}
\caption{Self-Excited Oscillations (color online). Panels (a) and (b) show the
reflected power at angular frequency $\omega_{\mathrm{p}}-\omega_{b}$ for
backward and forward frequency sweeps, respectively. The pump critical power
is $P_{\mathrm{c}}=-19.5$ dBm and the pump critical detuning is $\Delta
_{\mathrm{c}}=-2\pi\times0.7\operatorname{MHz}$. Panel (c) shows the reflected
power at angular frequency $\omega_{\mathrm{p}}$. Note the pulling in the
frequency response due to the cavity Kerr nonlinearity. The solid black,
dotted cyan and dotted green lines represent, respectively, supercritical
Hopf, subcritical Hopf and limit point of cycle bifurcations. The following
cavity parameters are employed for the bifurcation calculation $K_{a}%
/\omega_{a}=-1.44\times10^{-15}$ and $\gamma_{a3}/K_{a}$ $=0.04$.}%
\label{Sideband_results_fig}
%\end{center}
\end{figure*}

The microwave cavity is excited by injecting a monochromatic pump signal
having frequency $f_{\mathrm{p}}=\omega_{\mathrm{p}}/2\pi$ and amplitude
$b_{\mathrm{p}}$ into the feedline and monitoring the off-reflected signal
using either a spectrum analyzer or a diode connected to an oscilloscope [see
panel (a) of Fig. \ref{Setup_fig}]. The amplitude $b_{\mathrm{p}}$ is related
to the pump power $P_{\mathrm{p}}$ by $P_{\mathrm{p}}=\hbar\omega
_{a}\left\vert b_{\mathrm{p}}\right\vert ^{2}/2\gamma_{a1}$, where
$\gamma_{a1}$ represents the contribution to the total cavity linear damping
rate $\gamma_{a}$ due to cavity-feedline coupling \cite{Yurke_5054}. In the
absence of any optomechanical coupling (i.e. when the fiber is positioned far
from the trampoline) the cavity reflectivity exhibits bistability in a certain
region in the plane of pump parameters (frequency $f_{\mathrm{p}}$ and
amplitude $b_{\mathrm{p}}$) originates by cavity Kerr nonlinearity. The border
line of this region contains a cusp point, which is also known as the onset of
bistability point \cite{Yurke_5054}. The values of pump frequency and pump
amplitude at that critical point are labeled by $f_{\mathrm{c}}=\omega
_{\mathrm{c}}/2\pi$ and $b_{\mathrm{c}}$, respectively. In what follows we
employ normalized and dimensionless parameters for the pump detuning
$\Delta/\Delta_{\mathrm{c}}=\left(  \omega_{\mathrm{p}}-\omega_{a}\right)
/\left\vert \omega_{\mathrm{c}}-\omega_{a}\right\vert $ and for the pump
amplitude $b_{\mathrm{p}}/b_{\mathrm{c}}$.

\begin{figure*}

%\begin{center}
\includegraphics[width=\textwidth
%natheight=4.990000in,
%natwidth=9.781000in,
%height=3.3044in,
%width=6.4524in
]
{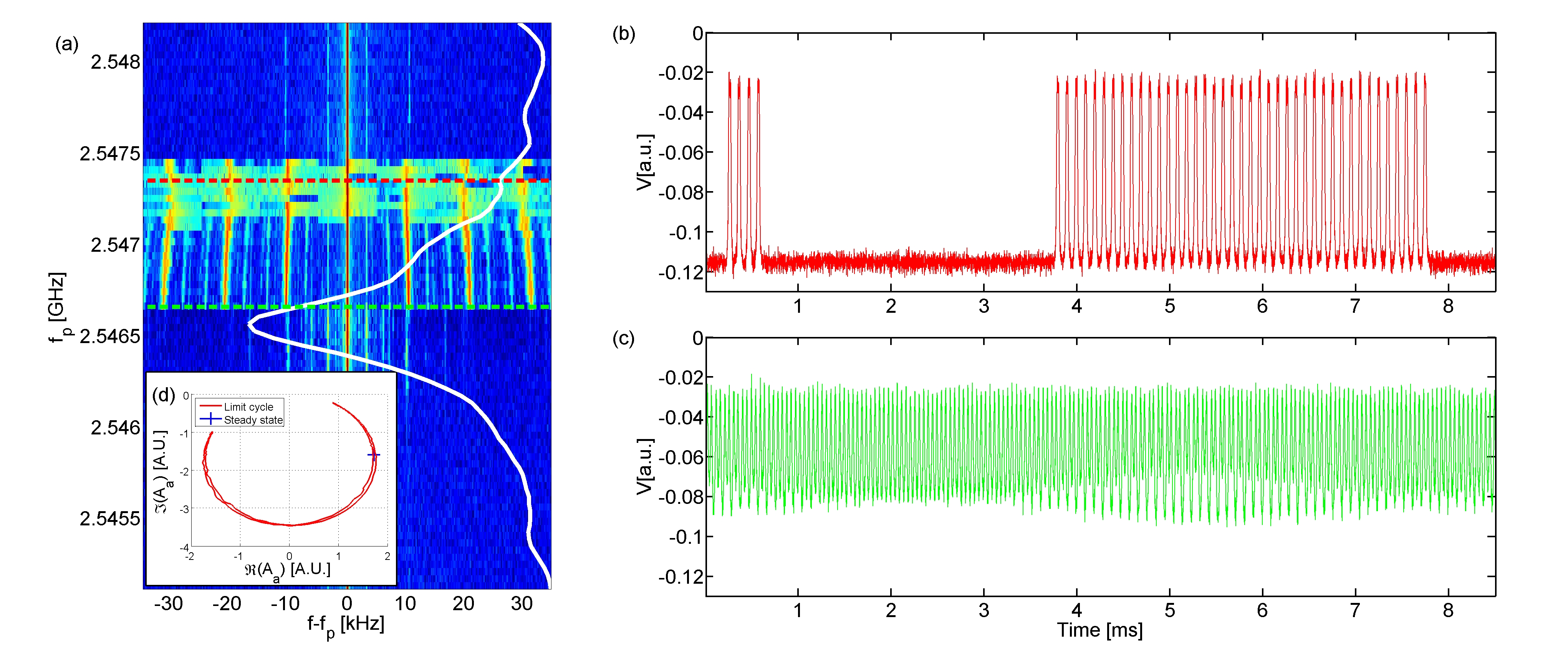}
\caption{Intermittency (color online). The measurement of the reflected signal
using a spectrum analyzer is seen in panel (a), whereas the time traces seen
in panels (b) and (c) are measured using an oscilloscope for two values of
$f_{\mathrm{p}}$, which are indicated in panel (a) by the two dotted lines.
The solid white line in panel (a) shows the reflected power at the pump
frequency $f_{p}.$ Noise-induced transitions between limit-cycle and steady
state are seen in panel (b). Panel (d) shows the projection of the limit cycle
(red line) and steady state (blue cross) on the complex $A_{a}~$ plane. }%
\label{intermittency_fig}%
%\end{center}
\end{figure*}

Panels (a) and (b) of Fig. \ref{Sideband_results_fig} show the reflected
microwave power, which is measured using a spectrum analyzer, at frequency
$f_{\mathrm{p}}-f_{b}$, where $f_{b}=\omega_{b}/2\pi$ is the mechanical
resonance frequency, for both forward and backward sweeps of the pump
frequency $f_{\mathrm{p}}$. A strong peak is found at frequency $f_{\mathrm{p}%
}-f_{b}$ [as well as at other harmonics $f_{\mathrm{p}}+nf_{b},$ where $n$ is
integer, as can be seen in panel (a) of Fig. \ref{intermittency_fig}] in a
certain region in the the plane of normalized pump parameters $\Delta
/\Delta_{\mathrm{c}}$ and $b_{\mathrm{p}}/b_{\mathrm{c}}$, inside which
self-excited oscillations occur. The height of the peak at frequency
$f_{\mathrm{p}}-f_{b}$ is plotted in panels (a) and (b) of Fig.
\ref{Sideband_results_fig}. The results obtained with backward sweep [panel
(a)] differ from those obtained with forward sweep [panel(b)], which indicates
that the cavity response is hysteretic due to bistability. Panel (c) depicts
the reflected power at $f_{\mathrm{p}}$.

Cavity nonlinearity plays a crucial role in the observed behavior of the
system. We employ the theoretical modeling of Refs.
\cite{Nation_104516,Buks_454} to account for cavity Kerr nonlinearity
\cite{Tholen_253509,Dahm_2002}. The equations of motion in the rotating frame
of the cavity for the annihilation operators $A_{a}$ and $A_{b}$ of the cavity
and mechanical resonator, respectively, are found to be given by
$\mathrm{d}A_{a}/\mathrm{d}t+\Theta_{a}=F_{a}$ and $\mathrm{d}A_{b}%
/\mathrm{d}t+\Theta_{b}=F_{b}$ where%
\begin{equation}
\Theta_{a}=\left[  i\Delta_{a}^{\mathrm{eff}}+\gamma_{a}+\left(  iK_{a}%
+\gamma_{a3}\right)  N_{a}\right]  A_{a}+b_{\mathrm{p}}\;,
\end{equation}%
\begin{equation}
\Theta_{b}=\left(  i\omega_{b}+\gamma_{b}\right)  A_{b}+i\Omega N_{a}\;,
\label{EOM_Ab}%
\end{equation}
and where $\Delta_{a}^{\mathrm{eff}}=\omega_{a}-\omega_{\mathrm{p}}%
+\Omega\left(  A_{b}+A_{b}^{\dag}\right)  $ is the effective cavity detuning,
$\Omega$ is the optomechanical coupling coefficient, $K_{a}$ is the cavity
Kerr nonlinearity constant, $\omega_{a}$ ($\omega_{b}$) is the cavity
(mechanical) angular resonance frequency, $\gamma_{a}$ ($\gamma_{b}$) is the
cavity (mechanical) linear damping rate, $\gamma_{a3}$ is the cavity nonlinear
damping rate, $N_{a}=A_{a}^{\dag}A_{a}$ is the cavity number operator and
$b_{\mathrm{p}}$ is the pump amplitude. The terms $F_{a}$ and $F_{b}$
represent white noise having (frequency independent) power spectrum given by
$S_{a}=2\Gamma_{a}\left(  1-e^{-\beta\hbar\omega_{a}}\right)  ^{-1}$ and
$S_{b}=2\gamma_{b}\left(  1-e^{-\beta\hbar\omega_{b}}\right)  ^{-1}$,
respectively, where $\Gamma_{a}=\gamma_{a}+2\gamma_{a3}\left\langle
N_{a}\right\rangle $, $\beta=1/k_{\mathrm{B}}T$, $k_{\mathrm{B}}$ is
Boltzmann's constant and $T$ is the temperature.

The stability map of the system is obtained using the numerical continuation
package MATCONT (URL: http://www.matcont.ugent.be/). First, a steady state
solution (i.e. solution to $\Theta_{a}=\Theta_{b}=0$) is found for each
operating point in the plane of pump parameters. Note that for the region seen
in Fig. \ref{Sideband_results_fig} the steady state is unique (since
$b_{\mathrm{p}}/b_{\mathrm{c}}<1$). Then MATCONT is employed to identify
bifurcations. The solid black, dotted cyan and dotted green lines in Fig.
\ref{Sideband_results_fig} represent, respectively, supercritical Hopf
(labeled as $\mathrm{H}_{-}$), subcritical Hopf (labeled as $\mathrm{H}_{+}$)
and limit point of cycle (LPC) bifurcations. In the numerical investigation
noise is disregarded and the operators $A_{a}$ and $A_{b}$ are treated as
c-numbers. The cavity parameters that are used for the numerical calculation
are listed in the caption of Fig. \ref{Sideband_results_fig}. The bifurcation
lines divide the region in the plane of pump parameters seen in Fig.
\ref{Sideband_results_fig} into three zones. In the zone between the
$\mathrm{H}_{-}$ and $\mathrm{H}_{+}$ bifurcations only a single limit-cycle
is found to be locally stable (though the existence of other locally stable
solutions cannot be ruled out), in the zone between the $\mathrm{H}_{+}$ and
LPC bifurcations bistability of a limit-cycle and a steady state occurs,
whereas elsewhere only a unique steady state is found.

Intermittency, i.e. random jumps between the limit-cycle and the steady state,
is experimentally observed for $b_{\mathrm{p}}/b_{\mathrm{c}}>0.2$. Fig.
\ref{intermittency_fig} shows frequency and time domain measurements taken
with normalized pump amplitude given by $b_{\mathrm{p}}/b_{\mathrm{c}}=0.58$.
Panel (a) shows the frequency decomposition of the reflected signal as the
pump frequency $f_{\mathrm{p}}$ is scanned and the white line shows the
reflected power at frequency $f_{\mathrm{p}}$. Regular self-excited
oscillations in the time domain can be seen in panel (c) for normalized
detuning $\Delta/\Delta_{\mathrm{c}}=0.2$. At larger detuning $\Delta
/\Delta_{\mathrm{c}}=1.5$ (in the bistable zone), however, random transitions
between the limit cycle and the steady state occur, as can be seen in panel
(b). The limit cycle and the steady state in the complex $A_{a}\ $projection
plane of phase space are plotted in panel (d). Note that the dynamics near the
steady state remains relatively slow even when it becomes locally unstable
(i.e. in the zone between the H$_{-}$ and H$_{+}$ bifurcations). Consequently,
in the presence of noise even a locally unstable steady state can give rise to
intermittency-like behavior provided that it is sufficiently close to the
limit cycle. Indeed, intermittency is experimentally observed on both sides of
the H$_{+}$ bifurcation.

In summary, we find that a subcritical Hopf bifurcation is the underlying
mechanism that is responsible for the experimentally observed intermittency.
While the current study is focused on the classical dynamics, future study
will explore the possibility of exploiting dynamical bistability for the
creation of macroscopic non-classical states of the optomechanical system
\cite{armour_440}.

This work was supported by the Israel Science Foundation, the bi-national
science foundation, the Deborah Foundation, the Israel Ministry of Science,
the Russell Berrie Nanotechnology Institute, the European STREP QNEMS Project
and MAFAT. The authors thank Ya'akov Schneider for helping in sample fabrication.

\bibliographystyle{apsrmp}
\bibliography{Eyal_Bib}

\end{document}